# Reductive Functionalization of Graphenides with Nickel(II) Porphyrin Diazonium Compounds

Hui-Lei Hou, Daniela Dasler, Frank Hauke, and Andreas Hirsch*[a]

[a] Dr. H.-L. Hou, D. Dasler, Dr. F. Hauke, Prof. Dr. A. Hirsch
Department of Chemistry and Pharmacy and Joint Institute of Advanced Materials and Processes (ZMP)
Friedrich-Alexander University of Erlangen-Nürnberg
Henkestrasse 42, 91054 Erlangen, Germany
E-mail: andreas.hirsch@fau.de

**Abstract:** A novel type of a graphene/porphyrin hybrid with a direct covalent linker between graphene and a nickel(II) tetraphenylporphyrin unit has successfully been synthesized by the reductive exfoliation/functionalization of potassium intercalated graphite as starting material. Moreover, we directly compared the reactivity of two structurally different porphyrin derivatives, namely, meso-NiTPP-$N_2^+$, with the diazonium group in 4'-position of a peripheral phenyl ring and a novel β-NiTPP-$N_2^+$ diazonium cation, where the diazonium group is directly attached to the β-pyrrolic position of the 18 π electron core. Due to sterical restraints, this intermediately generated porphyrin radical can not attack the extended π-surface of graphene and only a functionalization of present dangling bonds at the flake edges may be obtained in this case. All reaction products have been analyzed in detail be means of Raman- and UV/Vis spectroscopy as well as thermogravimetric analysis.

## Introduction

Graphene, a layered 2D and monoatomic thick carbon allotrope with $sp^2$-hybridized C atoms arranged in six-membered rings, has moved into the focus of physicists, materials engineers, and chemists within the last decade.[1-5] It's outstanding Young's modulus,[6] thermal conductivity,[7] mobility of charge carriers,[8] fracture strength[6] and extended surface area[9] renders graphene and especially the chemically modified analogs promising candidates for high-tech applications.[9],[10-11] One basic approach for the opening of a band gap[12-13],[14-15] and tailor made tuning of the electronic properties is its widely used covalent functionalization with diazonium compounds.[12] Here, highly reactive carbon centered radicals, obtained by an single electron transfer based extrusion of nitrogen, react with the $sp^2$ basal carbon atoms in graphene, leading to a covalent attachment of the respective functional entity.[16] This versatile approach has been utilized for the attachment of small aromatic molecules,[17-19] but examples for the direct binding of larger extended and redox-active π systems still remain scarce. With respect to novel solar cell-based applications some examples for graphene-porphyrin hybrids have been reported in literature.[20-25] These include graphene-TPP- (TPP = tetraphenylporphyrin) and graphene-PdTPP (PdTPP = palladium tetraphenylporphyrin) derivatives, which were synthesized *via* a 1,3-dipolar cycloaddtion reaction.[23] Another example is a zinc-porphyrin-triazole-graphene hybrid, obtained by a "click reaction" of a deprotected 4-(trimethylsilyl)ethynylaniline attached to reduced graphene oxide with the respective azide-terminated zinc-porphyrin component.[21] Chemically converted graphene (CCG)-porphyrin hybrids were first obtained *via* Suzuki reaction between iodophenyl-functionalized CCG and porphyrin boronic ester.[22] However, only few studies focused on the reaction of graphene and porphyrin diazonium compounds, which were formed *in situ* during the reaction.[25-26]

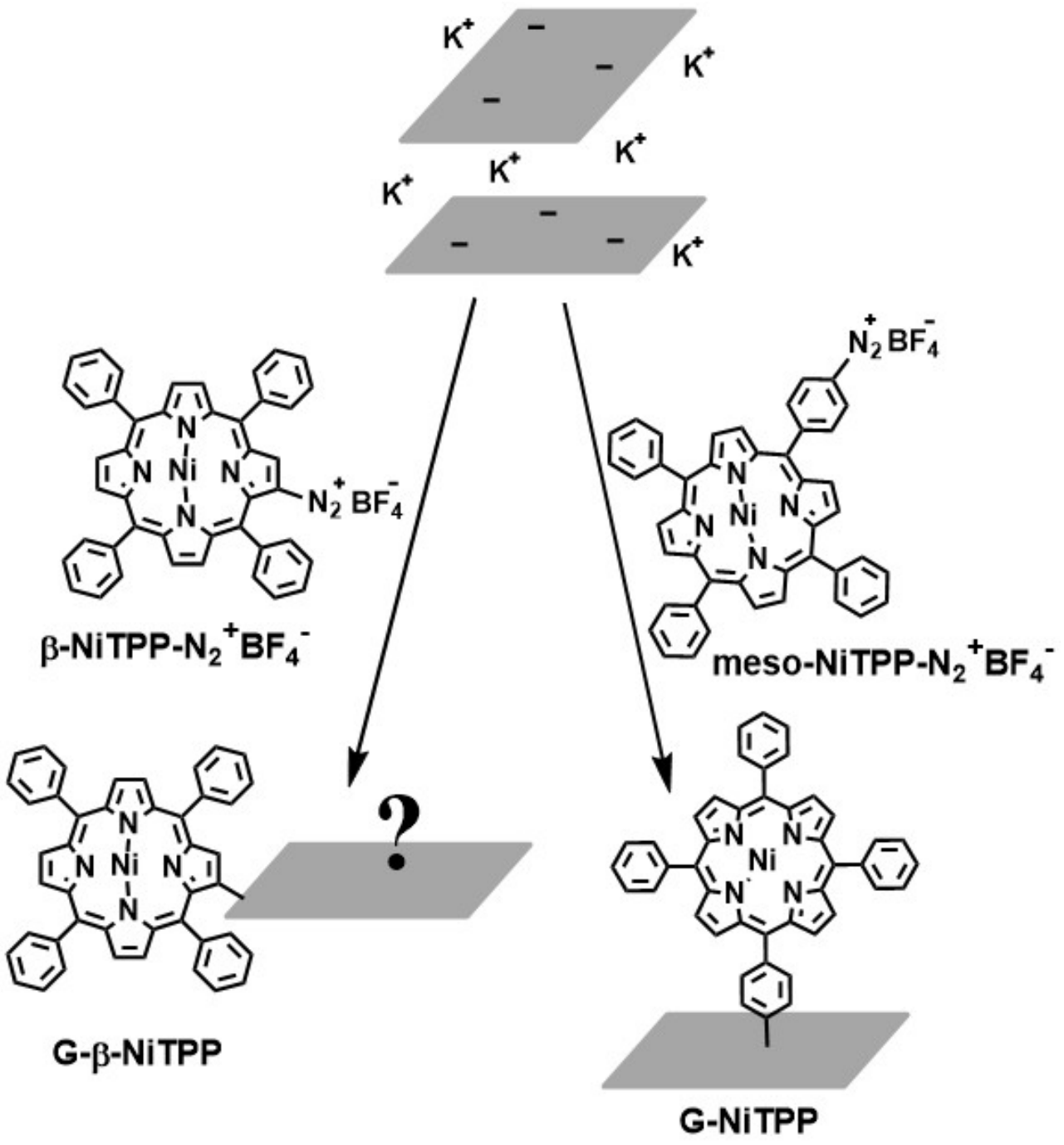

**Scheme 1.** Synthesis of graphene-porphyrin hybrids.

Nevertheless, the bulk one-pot synthesis of porphyrin-graphene hybrids on the basis of porphyrin diazonium salts in a reductive functionalization protocol utilizing graphite intercalation compounds (GICs) as starting materials has not been perused before. Herein, we report that negatively charged graphene – so called graphenides - obtained by an ultrasound based exfoliation of potassium intercalated graphite[27] ($KC_8$) with a molar ratio of K:C ratio of 1:8, can successfully undergo a direct, one-pot coupling with nickel-TPP diazonium salts (NiTPP) as trapping electrophiles. Moreover, we compared the reactivity and the regioselective impact of two differently substituted porphyrin systems, namely, meso-NiTPP-$N_2^+$ with the diazonium group in 4'-position of a peripheral phenyl ring and β-NiTPP-$N_2^+$ where the diazonium group is directly attached to the β-pyrrolic position of the 18 π electron core (Scheme 1).

## Results and Discussion

In order to evaluate the influence of the solvent on the reductive coupling of the two different NiTPP systems we used dried and degassed THF and for a direct comparison also DMF. Due to the high reactivity of the GICs and to prevent any side reactions, we carried out all functionalization reactions in an argon filled glovebox (oxygen content < 0.1 ppm, water content < 0.1 ppm). In this context, we used benzonitrile (PhCN) to quench any remaining negative charges[28] present in the intermediately generated graphenide/porphyrin derivatives - after the addition of the trapping electrophile and prior to the aqueous work up. All derivatives were characterized by means of Raman- and UV/Vis spectroscopy as well as by thermogravimetric analysis (TGA).

The initial UV/Vis spectroscopic investigation (Figure 1) of the finally obtained materials clearly exhibits a first indication for the successful formation of a graphene/porphyrin hybrid in the case of the reaction of $KC_8$ and the meso-NiTPP diazonium salt, both in DMF (G-NiTPP/DMF) as well as THF (G-NiTPP/THF).

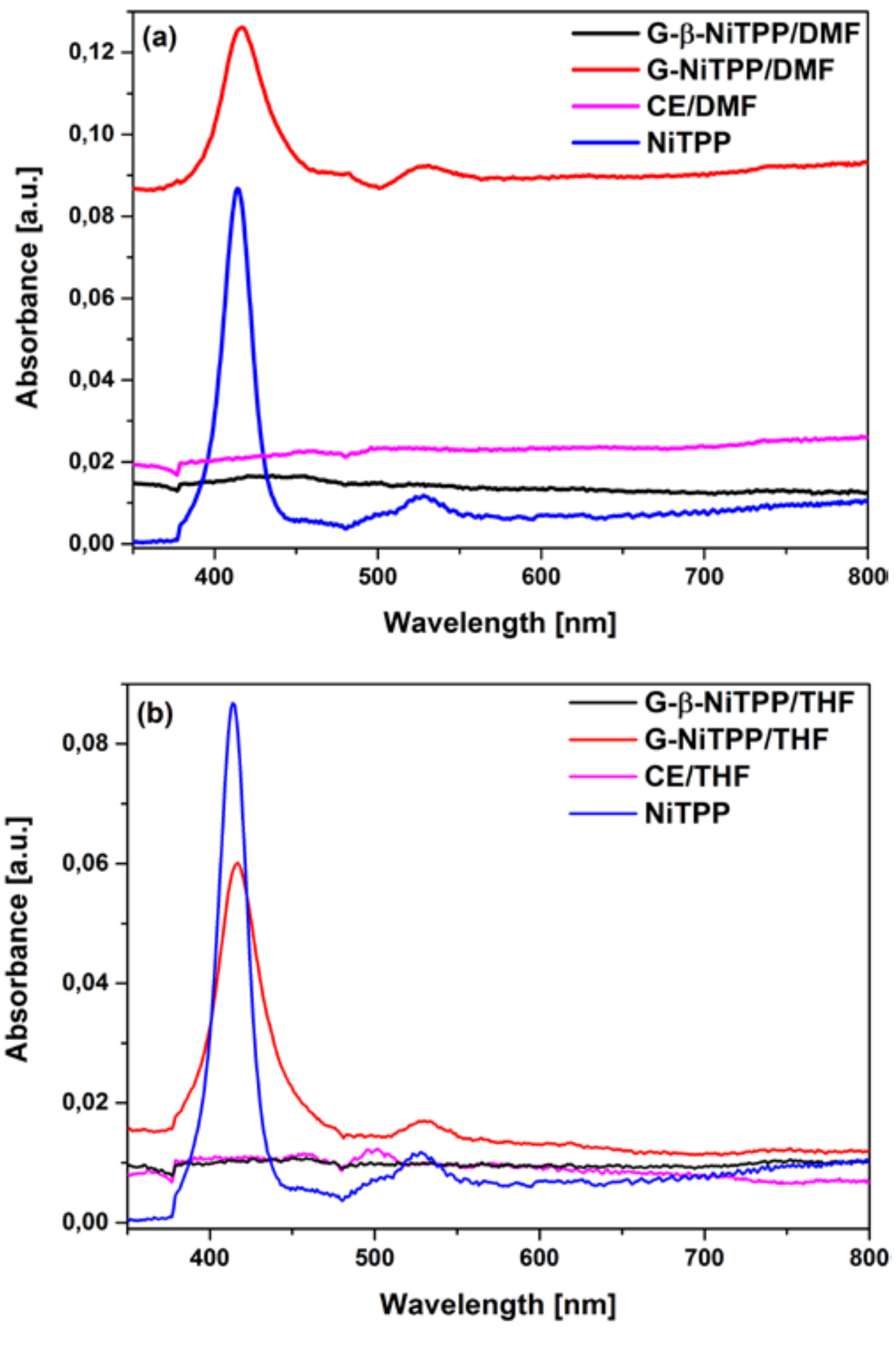

**Figure 1.** UV/Vis spectra of (a) G-β-NiTPP/DMF, G-NiTPP/DMF, NiTPP and CE/DMF and (b) G-β-NiTPP/THF, G-NiTPP/THF, NiTPP, and CE/THF in DMF.

Here, G-NiTPP/DMF and G-NiTPP/THF both display an characteristic and intense Soret band, centered at around 417 nm, and one weaker Q band at around 529 nm. The broadened and red-shifted (3 nm) main absorption band within the graphene/porphyrin hybrid, with respect to the spectroscopic feature of the pristine NiTPP (414 nm),[29-30] is indicative for a distinct electronic ground state coupling between graphene and the attached chromophoric unit.[22-23, 25]

Interestingly, in the obtained UV/Vis spectra of G-β-NiTPP/DMF and G-β-NiTPP/THF no evidence for the presence of the porphyrin system – physisorbed or chemisorbed - in the functionalized material can be found. This is corroborated by the spectral information obtained for two reference samples CE/DMF and CE/THF, were $KC_8$ was used as starting material and treated under exactly the same experimental conditions, except of the addition of any diazonium trapping electrophile. In the case of β-NiTPP-$N_2^+$ the direct covalent binding of the porphyrin unit *via* the formation of a β-pyrrole/$sp^3$ carbon bond would lead to a dramatic deformation of the porphyrin system due to the sterical demand of the attached phenyl substituents. This is in clear contrast to the situation present in meso-NiTPP-$N_2^+$. Here, the porphyrin is bound covalently perpendicular to the graphene basal plane *via* a peripheral phenyl ring, providing no sterical restraints and free rotational motion around the connecting $\sigma$ bond.

In order to obtain a better understanding of the geometrical situations in the G-β-NiTPP hybrid system we carried out a structure optimization at a PM3 level for β-NiTPP bound in the middle of the graphene basal plane and the configuration where the β-NiTPP unit is covalently bound the graphene edge, in the plain of the extended aromatic system (Figure 2).

As can be seen, in the case of the basal plane type (P/G-β-NiTPP, Figure 2a,c) the covalent attachment leads indeed to a pronounced deformation of the bound addend whereas in the edge type situation (E/G-β-NiTPP), a strain free situation can be obtained (Figure 2b,d), which is comparable to the situation found in the respective direct β-to-β coupled zinc porphyrin dimer.[31] Taking into account, that the ratio between basal plane carbon atoms in relation to edge carbon atoms is in clear disfavor for the latter ones, the amount of β-NiTPP bound in a edge type situation would be very low and probably under the UV/Vis detection limit. In this addition scenario the intermediately generated β-NiTPP would preferentially attack a peripheral dangling bond of the exfoliated graphene layer or abstract a terminating hydrogen atom, which would lead to the formation of unreactive NiTPP. This anticipation is supported by the recorded MALDI-TOF spectra of the washing eluent of the reaction outcome of β-NiTPP-$N_2^+$ with $KC_8$ in DMF (Figure S3) Here, NiTPP (m/z 670) is detected together with a distinct signal pattern for the respective radical-dimerization product $(NiTPP)_2$ (detected masses around m/z 1336). This is a clear indication that the intermediately generated β-NiTPP radicals, originating from a single electron transfer from the graphenides to the diazonium cation with subsequent $N_2$ extrusion, undergo a dimerization in competition to a rim proton abstraction process.

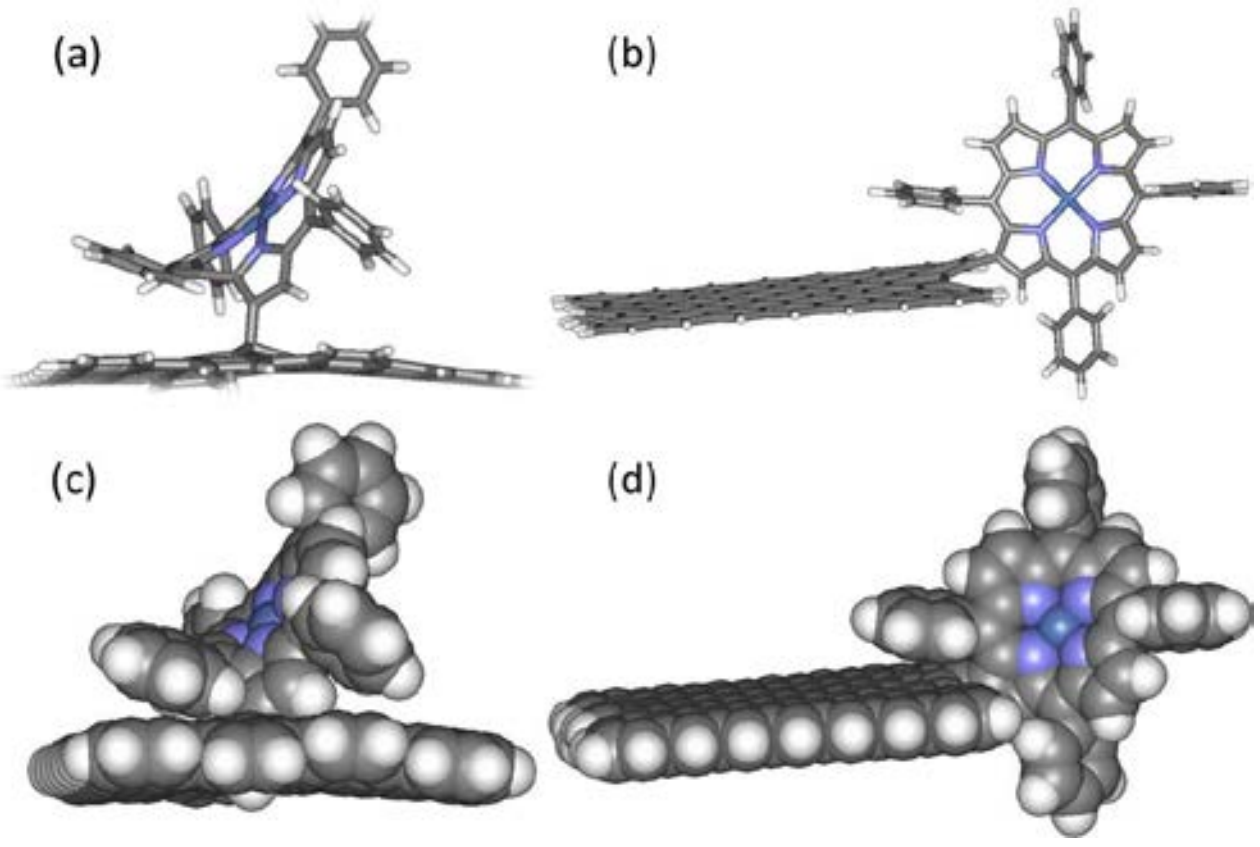

**Figure 2.** Calculated framework model configurations of (a) P/G-β-NiTPP and (b) E/G-β-NiTPP. Space-filling models of (c) P/G-β-NiTPP, and (d) E/G-β-NiTPP.

As stated above, in the case of the reductive exfoliation/functionalization of $KC_8$ with the meso-NiTPP-$N_2^+$ diazonium salt a covalently coupled porphyrin/graphene hybrid system (G-NiTPP) is easily formed due to a C-C bond formation between the 4'-position of a phenyl ring of NiTPP and one carbon atom of the graphene sheet. Our reductive bulk functionalization approach is mechanistically consistent with the literature reported derivatization of neutral graphene with intermediately generated phenyl radicals.[25] The analysis of the washing eluent of G-NiTPP/DMF (Figure S4) by MALDI-TOF corroborates that only a small amount of NiTPP dimmers are formed in the course of our porphyrin/graphene hybrid synthesis, as the intermediately generated meso-NiTPP radicals can easily react with the exfoliated graphene flakes. This is a fundamental difference to the situation found for the respective β-NiTPP radicals.

A clearer picture with respect to the basal plane alteration in the course of the reductive exfoliation/functionalization of GICs can be obtained by Raman spectroscopy. In general Raman spectroscopy and in particular, scanning Raman spectroscopy (SRS) represent a quantitative and reliable analytical tool for the bulk characterization of covalently functionalized carbon nanomaterials.[32-33] The Raman spectrum of graphene exhibits three main peaks - the defect induced D-band at 1,350 $cm^{-1}$, the G-band (in-plane stretching tangential mode) at 1,582 $cm^{-1}$, which is characteristic for the graphitic $sp^2$-carbon lattice, and the 2D-band at 2,700 $cm^{-1}$ providing information about the stacking or electronic decoupling of the graphene layers.[12, 34] The amount of $sp^3$ carbon atoms of the basal plane, bearing a covalently bound entity, correlates with the $I_D/I_G$ intensity ratio and therefore corresponds to the degree of functionalization. In combination with the frequency distribution of the individual $I_D/I_G$ values, which is a measure for the sample homogeneity, the averaged Raman spectra can be taken as a representative description of the respective bulk material.

The Raman spectra of G-NiTPP and G-β-NiTPP, obtained with THF and DMF as solvent, as well as the spectra of the respective reference samples are depicted in Figure 4.

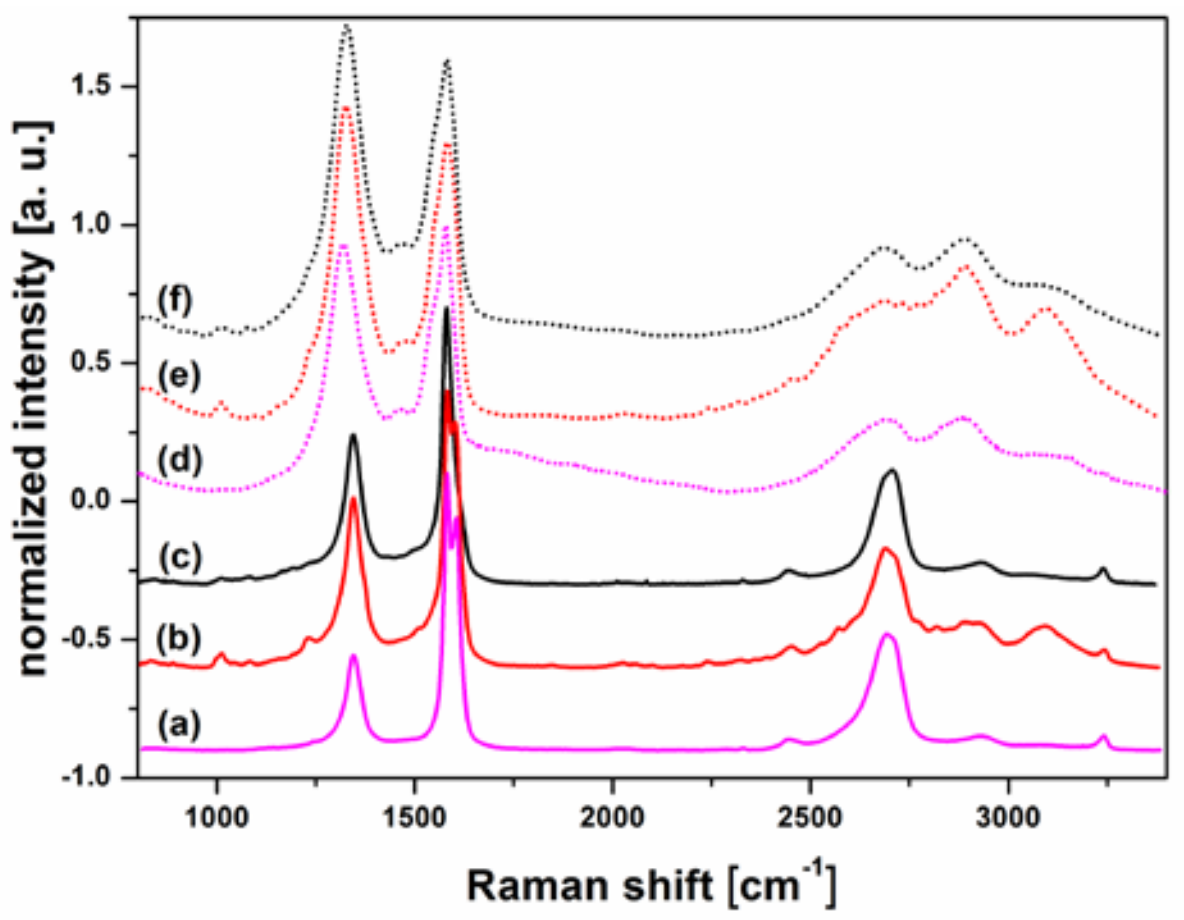

**Figure 4.** Raman spectra of (a) CE/DMF, (b) G-β-NiTPP/DMF, (c) G-NiTPP/DMF, (d) CE/THF, (e) G-β-NiTPP/THF, and (f). G-NiTPP/THF.

In the case of the samples obtained in DMF as solvent - G-β-NiTPP/DMF, G-NiTPP/DMF, and CE/DMF (Figure 4a-c) – a distinct increase of the D-band intensity is detected. The respective $I_D/I_G$ ratio of G-β-NiTPP/DMF (0.54) and G-NiTPP/DMF (0.61) is distinctively higher than that obtained for the reference CE/DMF (0.34). In the case of G-NiTPP/DMF the D-band intensity increase can easily be explained by the introduction of $sp^3$ basal carbon atoms covalently attached to the respective meso-NiTPP porphyrin unit.[35-37]. Remarkably, in the case of G-β-NiTPP/DMF also a pronounce D-band increase is detected and at the first glance counterintuitive to the discussion above. As G-β-NiTPP-$N_2^+$ only can attack the peripheral edges of the intermediately exfoliated graphene flakes, no D-band contribution of covalently attached porphyrin units in the carbon allotrope plane is expected. Nevertheless, as we have previously shown, graphenides represent highly reactive intermediates, which can undergo side reactions with any species present during their formation or in the course of the aqueous work up and therefore attribute to the detected D-band.[28, 38-39] On this basis, the detected $I_D/I_G$ intensity increase can be attributed to the attachment of proton- or hydroxyl functionalities.

It has to be mentioned, that the ratios of the respective $I_D/I_G$ values for both G-NiTPP/THF (1.1.) and G-β-NiTPP/THF (1.1), where THF has been used as solvent for the intermediate dispersion of the exfoliated graphenides, is remarkably higher than the respective values obtained in the case of DMF. This may be a first indication that DMF performs as the "more inert" solvent in the course of reductive exfoliation/Functionalization protocols of graphite intercalation compounds – a finding which has to be analyzed in detail in ongoing studies.

We also carried out temperature dependent Raman spectroscopy in order to follow in detail the thermal detachment of the covalently bound porphyrin unit in the case of G-NiTPP/DMF (Figure 5). As clearly can be seen, the Raman D-band intensity decreases during the heating experiment and reaches at 450 °C a value which is comparable to the situation found in the pristine starting graphite. This result suggests that G-NiTPP/DMF undergoes a thermally induced defunctionalization reaction by the loss of the covalently linked porphyrin units with a reversible restoration of its intact $sp^2$ basal plane.

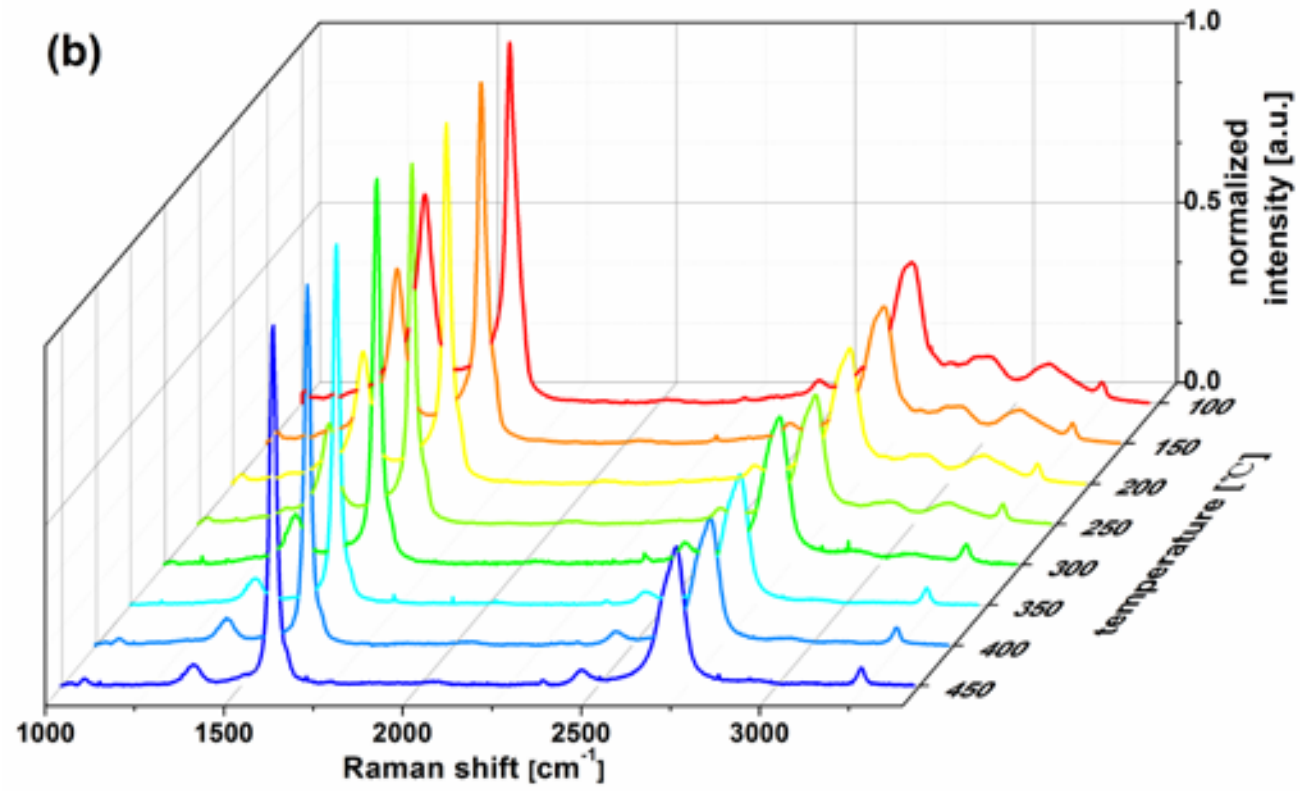

**Figure 5.** Temperature-dependent Raman spectra of G-NiTPP/DMF.

Interestingly, a distinct drop in the D-band intensity is observed for the temperature regime between 300 and 350 °C, and this should be attributed to cleavage of the σ bond, which connects the porphyrin unit to the graphene basal plane.[40] This process should in principle be mirrored in the respective data obtained by thermogravimetric analysis (TGA). Therefore, we carried out the respective characterization of the different porphyrin/graphene hybrids and the pristine NiTPP as a reference in a temperature regime between RT and 700 °C (Figure 6).

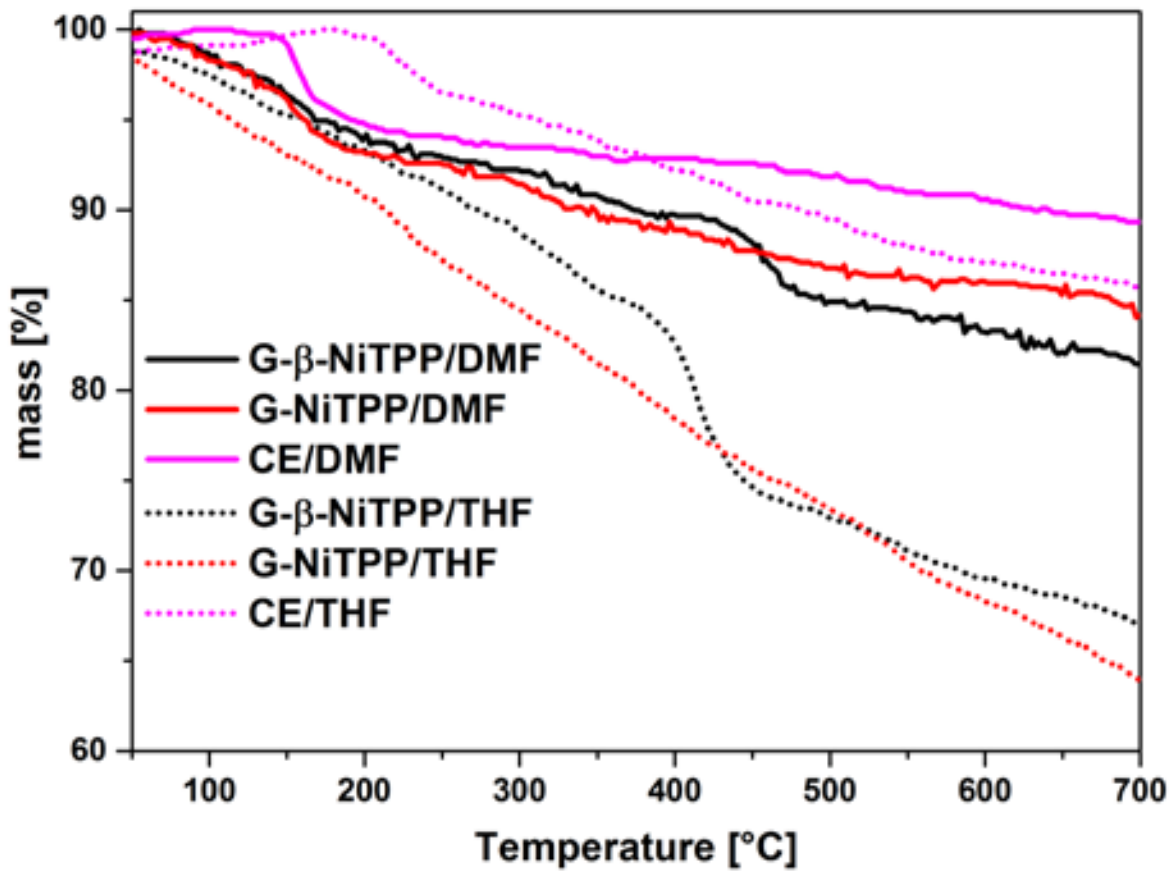

**Figure 6.** TGA curves of G-β-NiTPP, G-NiTPP, NiTPP, and of the control experiments under helium atmosphere.

For NiTPP (Figure S5) a distinct mass loss is detected between 400 and 500 °C which can be attributed to a thermal degradation of the porphyrin molecule. This trend is also observed for G-β-NiTPP/DMF and G-β-NiTPP/THF and may be attributed to intercalated and trapped porphyrin entities in the reaggregated bulk material.[18, 27] For the covalently bound graphene/porphyrin hybrids G-NiTPP/DMF and G-NiTPP no pronounced weight loss in the temperature regime between 300 °C and 350 °C, as suggested by the temperature dependent Raman investigations, can be detected by TGA. Nevertheless, if the total weight loss of these samples is compared to the weight loss obtained for the reference samples, CE/DMF and CE/THF, it becomes apparent that the additional weight loss can be attributed to covalently bound entities.

## Summary

In summary, G-NiTPP hybrids with direc covalent linkage have successfully been synthesized *via* a reductive one-pot bulk exfoliation/functionalization reaction of potassium intercalated graphite as starting material and a meso-NiTPP-$N_2^+$ diazonium salt. A novel, structurally different porphyrin diazonium compound, where the diazonium leaving group is directly attached to the β-pyrrolic position of the 18 π electron core, has been synthesized and utilized as a potential test component in order to screen basal plan functionalization *vs.* graphene rim functionalization. On the basis of PM3 structure optimizations it can be anticipated that this type of electrophilic trapping reagent can not attach to carbon atoms in the center of the extended π-electron surface of graphene. We have found that it is difficult to unambiguously judge whether these β-NiTPP derivatives are covalently attached on the edges of graphene sheet or if they simply undergoes a deactivation in the course of a dimerization process. Therefore, ongoing experiments are needed in order to answer the fundamental question of plane *vs.* edge functionalization in graphene chemistry.

## Acknowledgement

The authors thank the Deutsche Forschungsgemeinschaft (DFG-SFB 953, Project A1 “Synthetic Carbon Allotropes”) for financial support. The research leading to these results has received partial funding from the European Union Seventh Framework Programme under grant agreement no.604391 Graphene Flagship. H.-L. H. acknowledges the support from the Alexander von Humboldt-Foundation.

Supporting Information

# Reductive Functionalization of Graphenides with Nickel(II) Porphyrin Diazonium Compounds

Hui-Lei Hou, Daniela Dasler, Frank Hauke, and Andreas Hirsch*

## 1. Materials

Synthetic spherical graphite (SGN18, 99.99 % C, TGA residue 0.01 % wt – Future Carbon, Germany) with a mean grain size of 18 µm and a specific surface area of 6.2 $m^2$/g was used after annealing under vacuum (300 °C). Chemicals and solvents were purchased from Sigma Aldrich Co. (Germany) and were used as-received if not stated otherwise. DMF, THF and PhCN were dried by 3 Å molecular sieves to remove water. The determined residual water content by Karl Fischer titration was below 5 ppm. Oxygen was removed by pump freeze treatment (six iterative steps).

## 2. Equipment and Characterization

**Glove Box:** Sample synthesis and preparation was carried out in an argon filled LABmasterpro sp glove box (MBraun), equipped with a gas purifier and solvent vapor removal unit: oxygen content < 0.1 ppm, water content < 0.1 ppm.

**Raman Spectroscopy:** The Raman spectroscopic characterization was carried out on a Horiba Jobin Yvon LabRAM Aramis. A laser (Olympus LMPlanFl50x, NA 0.50) with an excitation wavelength of 532 nm and a spot size of ~ 1 µm was used. The spectrometer was calibrated in frequency using crystalline graphite before the measurement. The measurement data were obtained through a motorized x-y table in a continuous line scan mode (SWIFT-module).

The temperature dependent Raman measurements were performed in a Linkam stage THMS 600, equipped with a liquid nitrogen pump TMS94 for temperature stabilization under a constant flow of nitrogen. The measurements were carried out on Si/$SiO_2$ wafers with a heating rate of 10 K·$min^{-1}$.

LabSpec 6 was used for the recording of the spectra. Subsequently, the obtained spectroscopic information was analyzed after export into Origin 9.0G as 2D data matrices. All following operations (e. g. statistic treatment, graphics) were carried out within this software.

**Thermogravitmetric analysis coupled with mass spectrometry (TGA/MS):** The thermogravimetric analysis was performed on a Netzsch STA 409 CD equipped with a Skimmer QMS 422 mass spectrometer (MS/EI). The measurements were implemented from 25-700 °C with a 10 K/min gradient. Subsequently, the system was cooled down to room temperature. The initial sample weights were about 4 mg, and the whole experiment was executed under helium atmosphere with a flow of 80 mL/min.

**UV-vis spectroscopy:** The spectra were recorded on a Perkin Elmer Lambda 1050 spectrometer in DMF at room temperature in quartz cuvettes with 1 cm path length. The baseline was measured against air.

**$^{1}$H- and $^{13}$C-NMR spectroscopy:** $^{1}$H- and $^{13}$C-NMR measurements were carried out on a BRUKER Advance 400 and JEOL JNM EX 400 instrument at 400 MHz and 100 MHz in $CDCl_3$ at room temperature. All data were analyzed with MestReNova LITE.

**Mass spectrometry (MS):** All mass spectra were recorded on a SHIMADZU BIOTECH AXIMA Confidence in MALDI-TOF mode without a matrix or with 2,5-dihydroxybenzoic acid (DHB), or trans-2-[3-(4-tertbutylphenyl)2-methyl-2-propenylidene]malononitrile (DCTB).

**Infrared spectroscopy (IR):** All spectra were measured on a BRUKER TENSOR 27 in ATR mode (diamond) directly in solid substance. The obtained data were analyzed in Origin 9.0G.

## 3. Experimental Details

**Synthesis of (5,10,15,20-tetraphenylporphyrinato) nickel(II) (NiTPP):** TPP (1.0 g, 1.6 mmol) was dissolved in 200 mL DMF and then $Ni(AcO)_2 \cdot 4H_2O$ (7.9 g, 32 mmol) was added. The mixture was stirred for 6 h under reflux conditions. Then 200 mL cold water was added to quench the reaction. The mixture was filtered and washed with methanol, yielding 1.0 g (90%) of NiTPP.

**Synthesis of (2-amino-5,10,15,20-tetraphenylporphyrinato) nickel(II) (β-NiTPP-NH$_2$):** NiTPP (500 mg, 0.8 mmol) was dissolved in 325 mL $CHCl_3$ and then $Cu(NO_3)_2 \cdot 3H_2O$ (600 mg, 2.5 mmol) and acetic anhydride (36 mL) were added. The mixture was stirred for 3 h under reflux conditions. Then 100 mL of 0.1 M NaOH solution were added, the organic layer was separated and washed with water before being dried over anhydrous $Na_2SO_4$. Finally, the residue was filtered and the remaining solvent traces were removed under vacuum. The residue was chromatographed on a silica gel column eluting with DCM/hexane (v/v = 5:5) to afford β-nitro NiTPP (β-NiTPP-NO$_2$). The resulting NiTPP-NO$_2$ was dissolved in $CHCl_3$ (40 mL), and then tin powder (5 g) and concentrated HCl (20 mL) were added. The reaction mixture was stirred at 40 °C for 3 hour. The mixture was filtered and the filtrate was extracted with DCM. The organic layer was combined and was washed once with water, passed through a plug of dry silica gel, and concentrated in vacuum. The residue was recrystallized from DCM/MeOH to afford 170 mg (overall yield is 31%) β-NiTPP-$NH_2$ (see Scheme 2).

Spectral characterization of β-NiTPP-$NH_2$: positive MS, m/z calculated for $C_{44}H_{29}N_5Ni^+$ $[M]^+$ 685, found 685. $^1H$ NMR (400 MHz, $CDCl_3$) δ 4.22 (s, 2H, β-NH2), 7.62-7.73 (m, 13H, 12Ph-H + 1β-H), 7.95-8.01 (m, 8H, Ph-H), 8.56 (d, *J* = 4 Hz, 1H, β-H), 8.62-8.70 (m, 5H, β-H) ppm. UV-vis ($CH_2Cl_2$) $\lambda_{max}$ 412, 535, 590 nm.

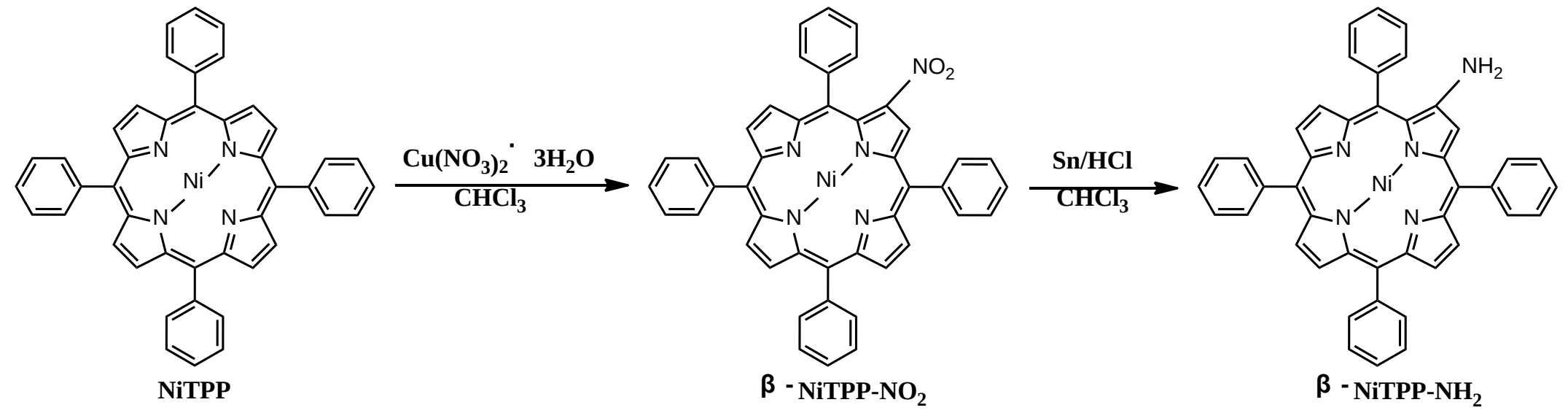

Scheme S1. Synthesis of β-NiTPP-$NH_2$ compound

**Synthesis of 5-(4-Aminophenyl)-10,15,20-triphenylporphyrin (meso-TPP-$NH_2$)**: Mesotetraphenylporphyrin (TPP, 1.0 g, 1.6 mmol) was dissolved in trifluoroacetic acid (TFA, 50 mL) and then sodium nitrite (200 mg, 2.9 mmol) was added. After 3 min stirring at room temperature, the reaction mixture was poured into water (300 mL) and extracted with dichloromethane (DCM, 6 × 50 mL). The organic layer was washed once with saturated aqueous $NaHCO_3$ and water, respectively, and subsequently dried over anhydrous $Na_2SO_4$. Finally, the residue was filtered and the remaining solvent traces were removed under vacuum. The residue was chromatographed on a silica gel column eluting with DCM/hexane (v/v = 5:5) to afford nitro TPP (TPP-$NO_2$). TPP-$NO_2$ (250 mg, 0.38 mmol) was dissolved in concentrated hydrochloric acid (10 mL) and, while stirring, tin(II) chloride (262 mg, 1.2 mmol) was carefully added at 67 °C for 2 h under $N_2$ before being poured into cold water (30 mL). The aqueous solution was neutralized with aqueous ammonium hydroxide to pH = 8. Subsequently, the solution was extracted with DCM until colorless. The organic layer was then concentrated under vacuum and the residue was chromatographed on a silica gel column eluting with DCM. The final residue was recrystallized from methanol, yielding 192 mg (80%) of amino porphyrin TPP-$NH_2$.

Spectral characterization of TPP-$NH_2$: positive MS, m/z calculated for $C_{44}H_{31}N_5^+$ $[M]^+$ 630, found 630. $^1H$ NMR (400 MHz, $CDCl_3$) δ -2.74 (br, 2H),

3.91 (s, 2H), 6.99 (d, *J* = 8 Hz, 2H), 7.72-7.75 (m, 9H), 7.97 (d, *J* = 8 Hz, 2H), 8.20-8.23 (m, 6H), 8.83-8.84 (m, 6H), 8.94 (d, *J* = 4 Hz, 2H) ppm. UV-vis ($CH_2Cl_2$) $\lambda_{max}$ 419, 517, 554, 593, 648 nm.

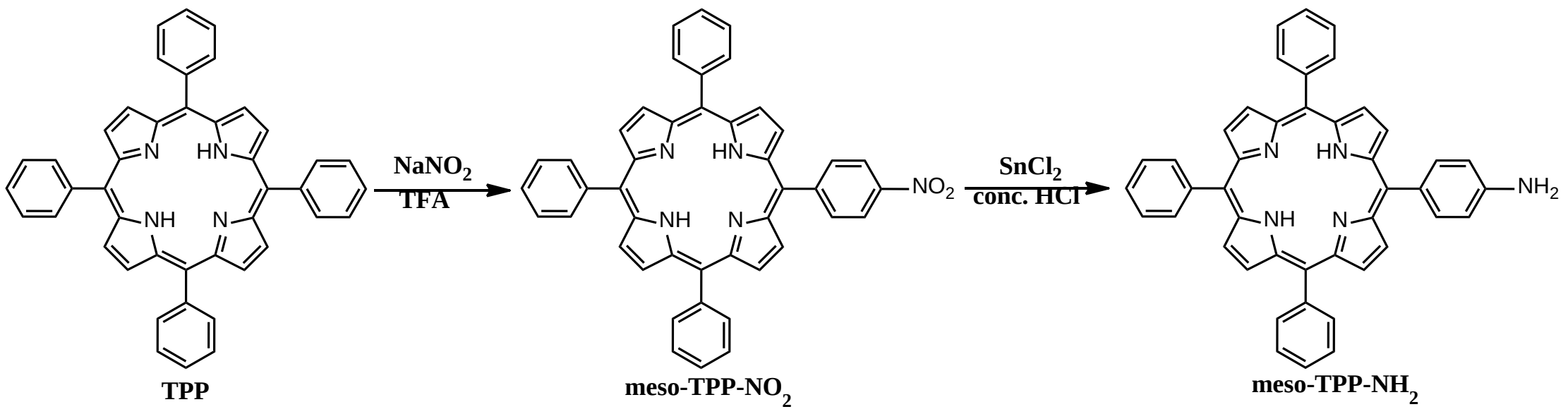

Scheme S2. Synthesis of meso-TPP-NH2 compound

**Synthesis of (5-(4-Aminophenyl)-10,15,20-triphenylporphyrinato) nickel(II) (meso-NiTPP-NH2):** Meso-TPP-NH2 (100 mg, 0.16 mmol) was dissolved in DMF (100 mL) and then nickel(II) acetate tetrahydrate (796 mg, 3.2 mmol) was added. After 6 h stirring under reflux, the reaction mixture was cooled down to room temperature and poured into water (100 mL). Subsequently, the mixture was filtered and the residue was chromatographed on a silica gel column eluting with DCM to afford 97 mg (88%) meso-NiTPP-NH2.

Spectral characterization of meso-NiTPP-NH2: positive MS, m/z calculated for $C_{44}H_{29}N_5Ni^+$ $[M]^+$ 685, found 685. $^1H$ NMR (400 MHz, $CDCl_3$) δ 3.93 (broad, 2H, $NH_2$), 6.96 (d, *J* = 8 Hz, 2H), 7.65-7.69 (m, 9H), 7.78 (d, *J* = 8 Hz, 2H), 8.00-8.02 (m, 6H), 8.73-8.74 (m, 6H), 8.84 (d, *J* = 4 Hz, 2H) ppm. UV-vis ($CH_2Cl_2$) $\lambda_{max}$ 417, 529 nm.

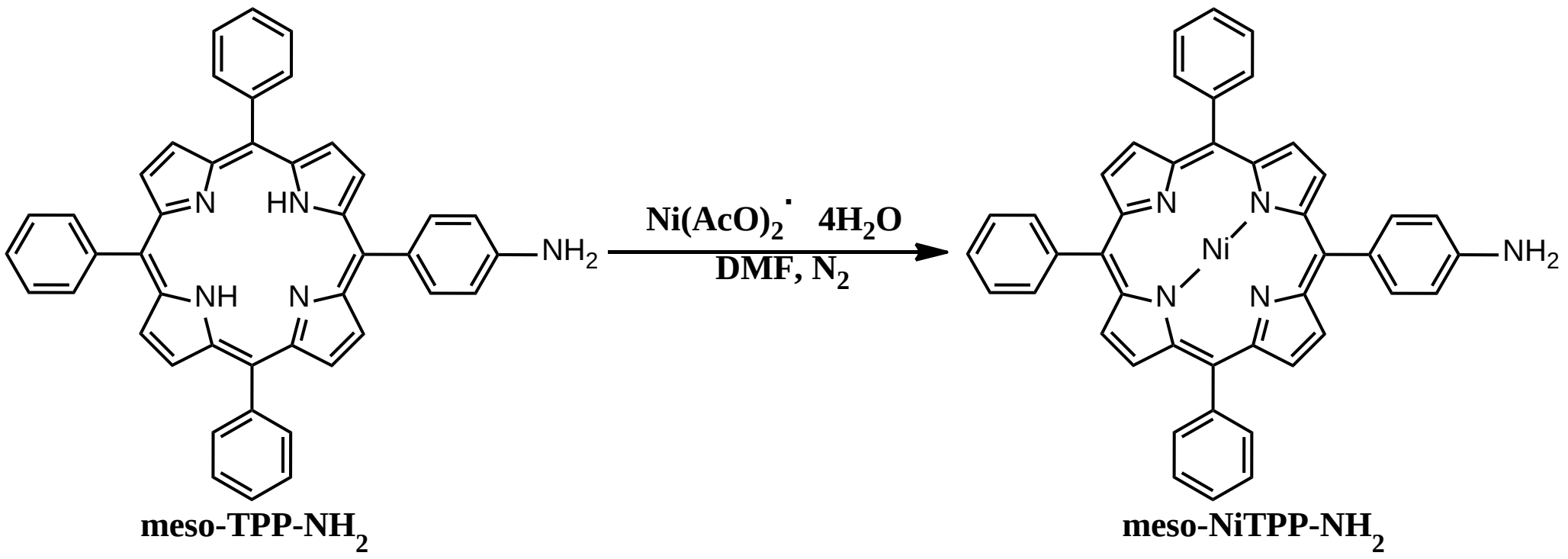

Scheme S3. Synthesis of meso-NiTPP-NH2 compound

**Diazotation of (2-amino-5,10,15,20-tetrapheny1porphyrinato) nickel(II) to afford β-NiTPP-$N_2^+BF_4^-$:** Firstly, β-NiTPP-NH2 (165 mg, 0.24 mmol, 1 eq.) was added into $HBF_4$ (0.1 mL in 48 wt% $H_2O$, 1.59 mmol, 6.63 eq.). Subsequently, 10 mL acetic acid was added and isoamylnitrite (0.1 mL, 0.75 mmol, 3.13 eq.) dissolved in 5 mL acetic acid was slowly dropped to the solution. After 15 minutes, the reaction mixture was quenched with 10 mL diethyl ether and stored at -22 °C overnight. The green solid was filtered off over a 0.2 μm pore filter and washed with 5 mL diethyl ether. Compound β-NiTPP-$N_2^+BF_4^-$ was obtained with an isolated yield of 63% (118 mg).

The ATR-IR spectrum of β-NiTPP-$N_2^+BF_4^-$ exhibits the characteristic band of diazonium salt at around 2170 $cm^{-1}$.

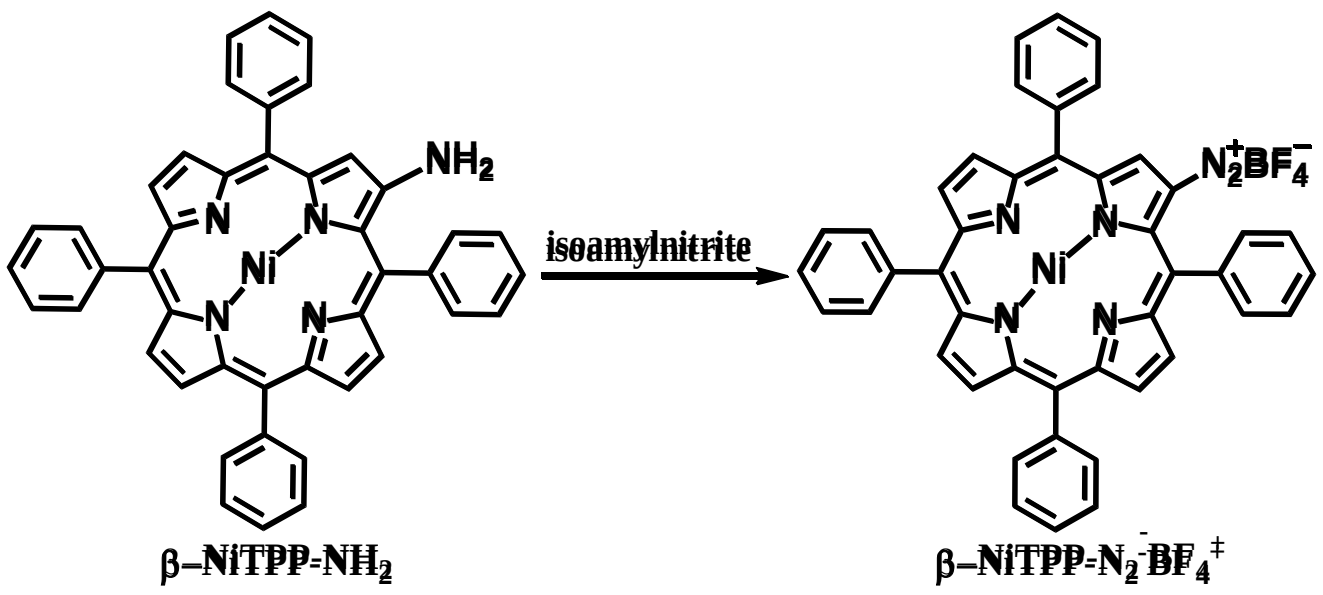

Scheme S4. Synthesis of β-NiTPP-$N_2^+BF_4^-$ compound

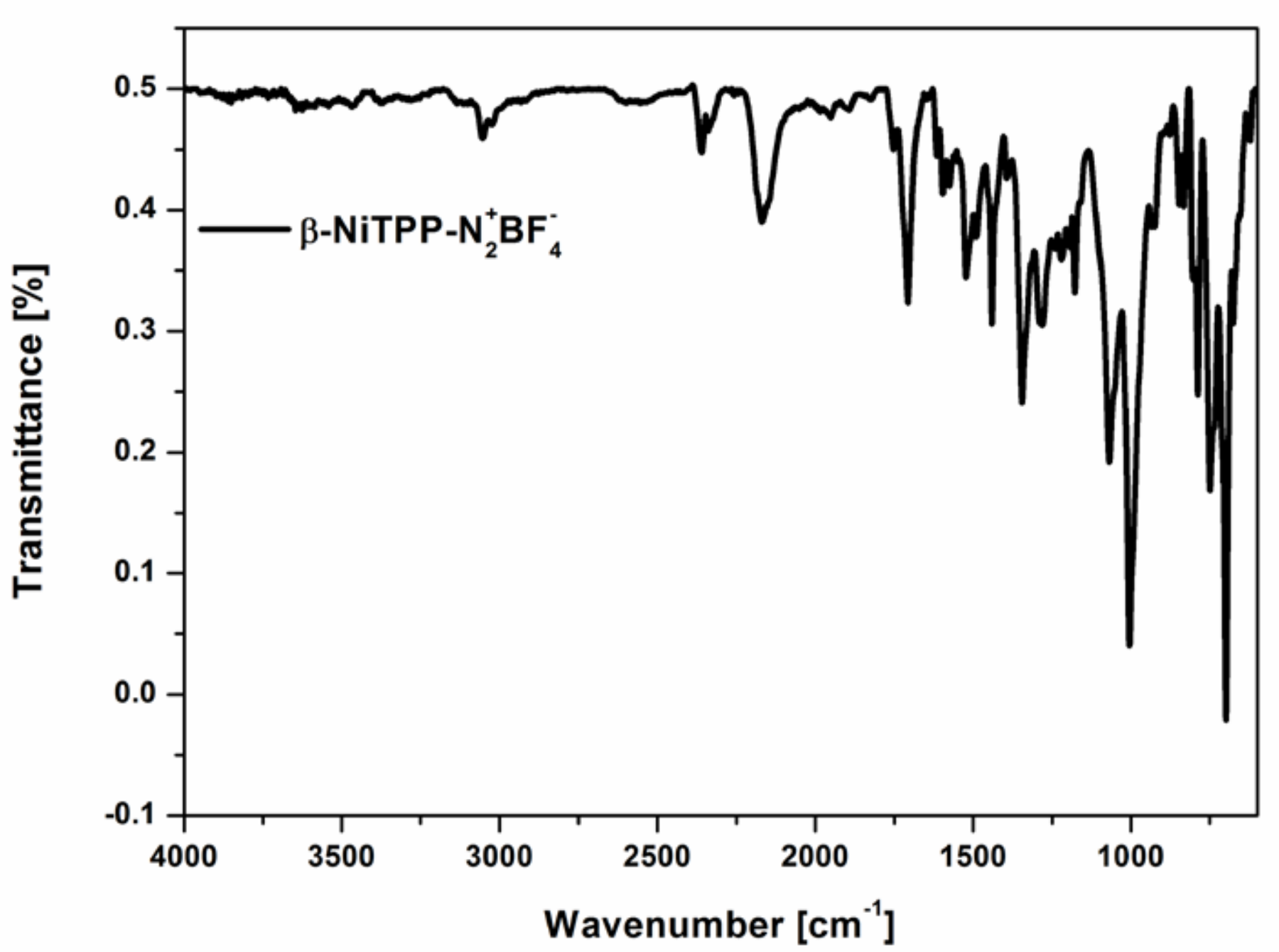

Figure S1. The ATR-IR spectrum of β-NiTPP-$N_2^+BF_4^-$

**Diazotation of (5-(4-Aminophenyl)-10,15,20-triphenylporphyrinato)-nickel(II) to afford meso-NiTPP-$N_2^+BF_4^-$:** The preparation is similar to the synthesis of β-NiTPP-$N_2^+BF_4^-$, except meso-NiTPP-$NH_2$ was used instead of β-NiTPP-$NH_2$. Compound meso-NiTPP-$N_2^+BF_4^-$ was obtained with an isolated yield of 77% (145 mg).

The IR spectrum of meso-NiTPP-$N_2^+BF_4^-$ exhibits the characteristic band of diazonium salt at around 2280 $cm^{-1}$.

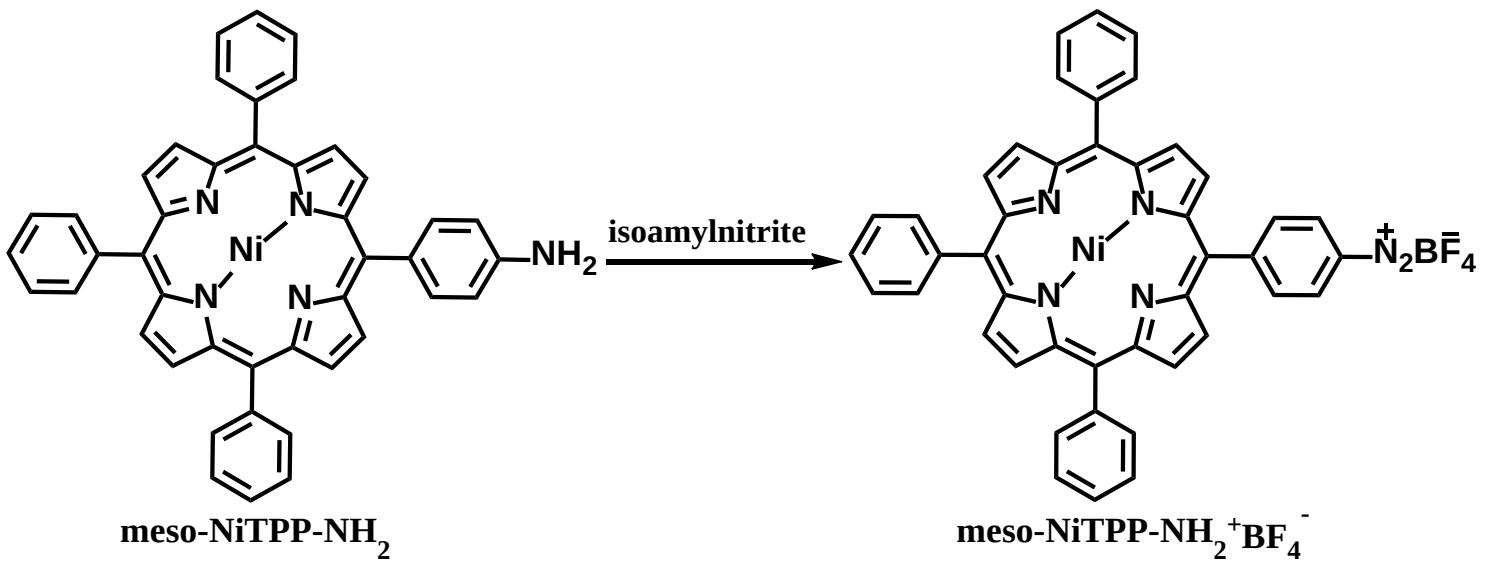

Scheme S5. Synthesis of meso-NiTPP-$N_2^+BF_4^-$ compound

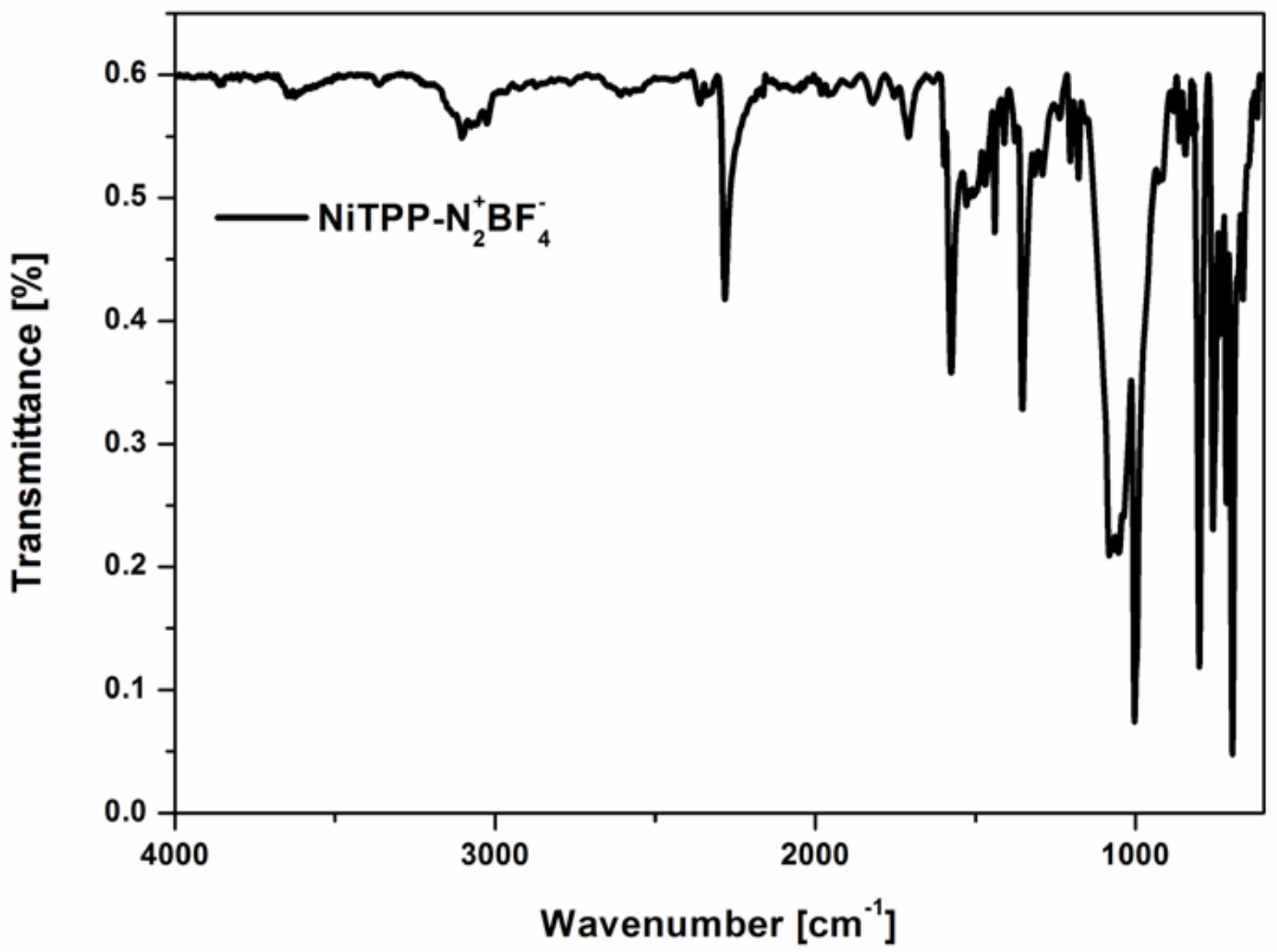

Figure S2. The ATR-IR spectrum of NiTPP-$N_2^+BF_4^-$

**Synthesis of potassium intercalation graphite ($KC_8$):** Pristine spherical graphite (SGN18) was heated under vacuum and transferred into an Ar atmosphere in the glovebox. Then 72 mg (1 mmol C) graphite and 30 mg (0.125 mmol K) potassium were mixed and heated at 100 °C overnight in order to obtain the brown golden $KC_8$ graphite potassium intercalation compound.

**Synthesis of G-β-NiTPP/DMF and G-β-NiTPP/THF (NiTPP = (tetraphenylporphyrinato)nickel(II)):** Firstly, $KC_8$ (17 mg, 1 mmol, 1 eq.) was dissolved in 20 mL of dried and degassed THF or DMF, respectively, at room temperature *via* 2 min sonication (sonotrode: Amp 50 %, ultrasonic exposure time 0.5 s, 1 s interval). The reaction mixture was then exported from the glovebox. Subsequently, 0.125 mmol (0.125 eq.) diazonium salt β-NiTPP-$N_2^+BF_4^-$ were added under Ar atmosphere. After 20 h stirring, the reaction was quenched by the addition of 10 mL PhCN under Ar flow. 50 mL cyclohexane was added to the solution, which was then washed three times with 50 mL water. The organic phase was filtered over a reinforced cellulose membrane filter (0.2 µm pore). The grey solid was washed with acetone, THF, and toluene.

**Synthesis of G-NiTPP/DMF and G-NiTPP/THF (NiTPP = (tetraphenylporphyrinato)nickel(II)):** The procedure was similar to those for preparation of G-β-NiTPP, except NiTPP-$N_2^+BF_4^-$ was used instead of β-NiTPP-$N_2^+BF_4^-$.

**Control experiments (CE/DMF and CE/THF):** The procedure was similar to those for preparation of G-β-NiTPP, but without addition of any porphyrin diazonium salts.

**4. Structure optimization:** The geometry optimization of the different porphyrin/graphene hybrids was performed using the Spartan'10 program (Wavefunction Inc.) at PM3 level.

## 5. Further characterization and spectroscopic information

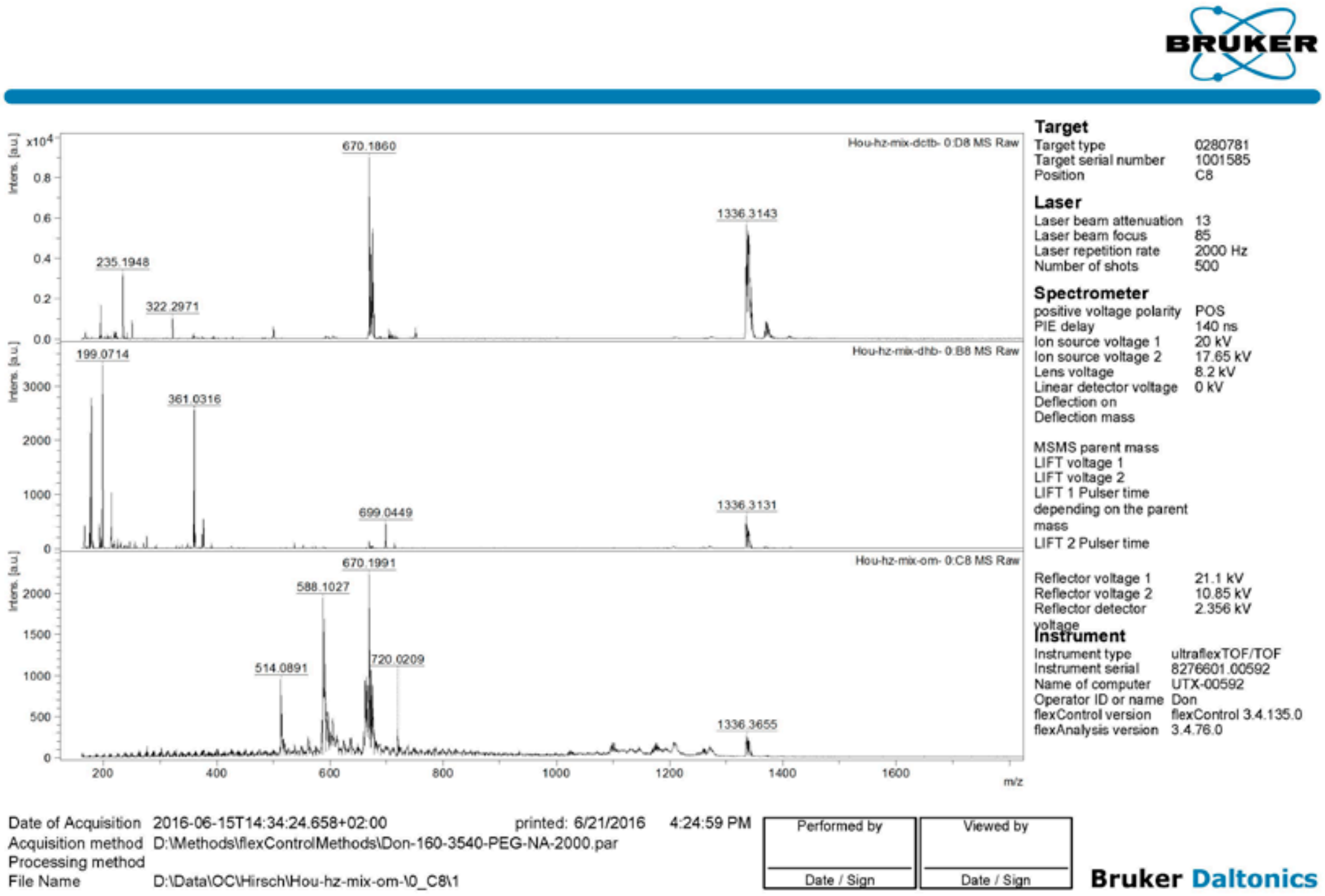

Figure S3. MALDI-TOF spectrum of G-β-NiTPP washing eluent.

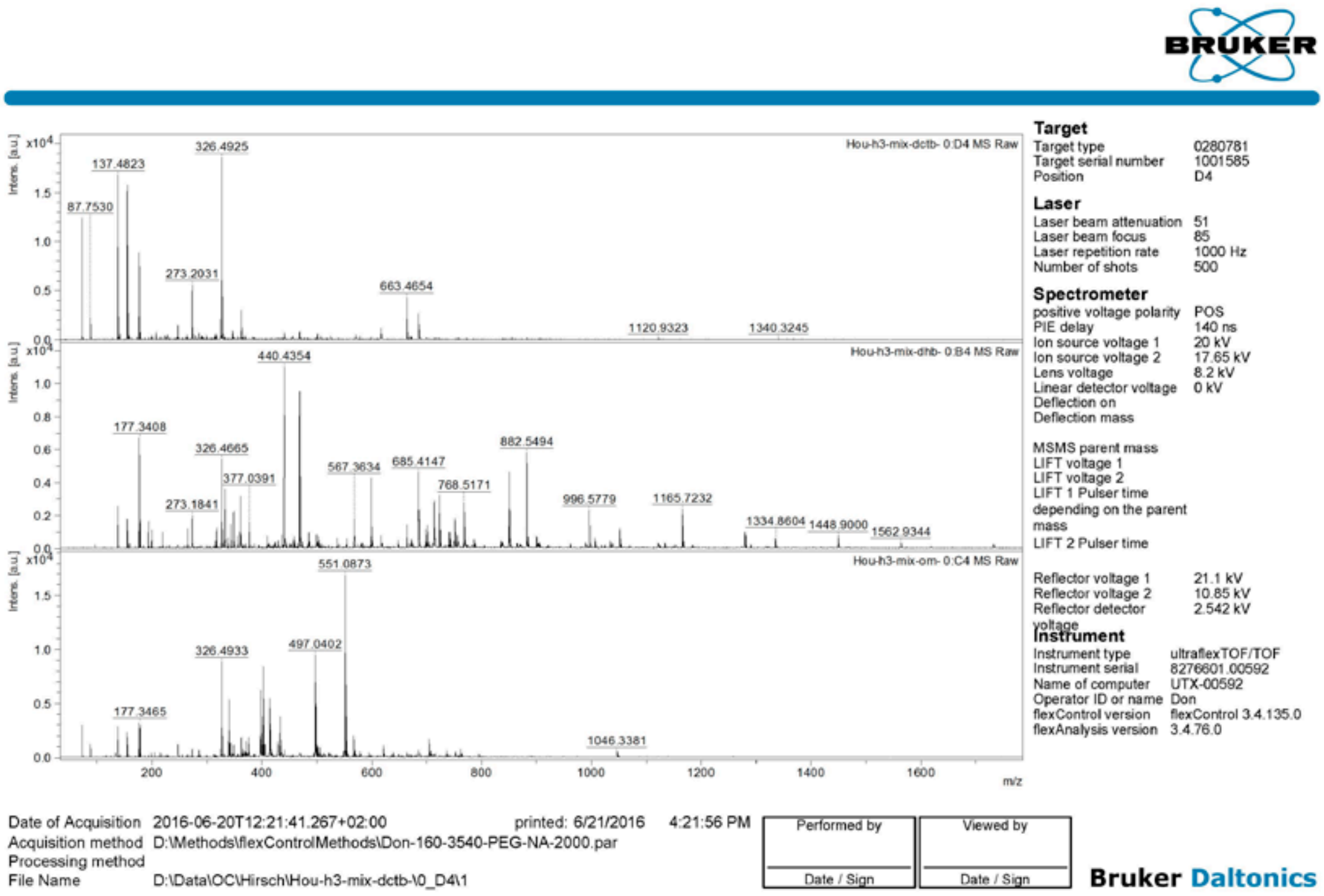

Figure S4. MALDI-TOF spectrum of G-NiTPP washing eluent.

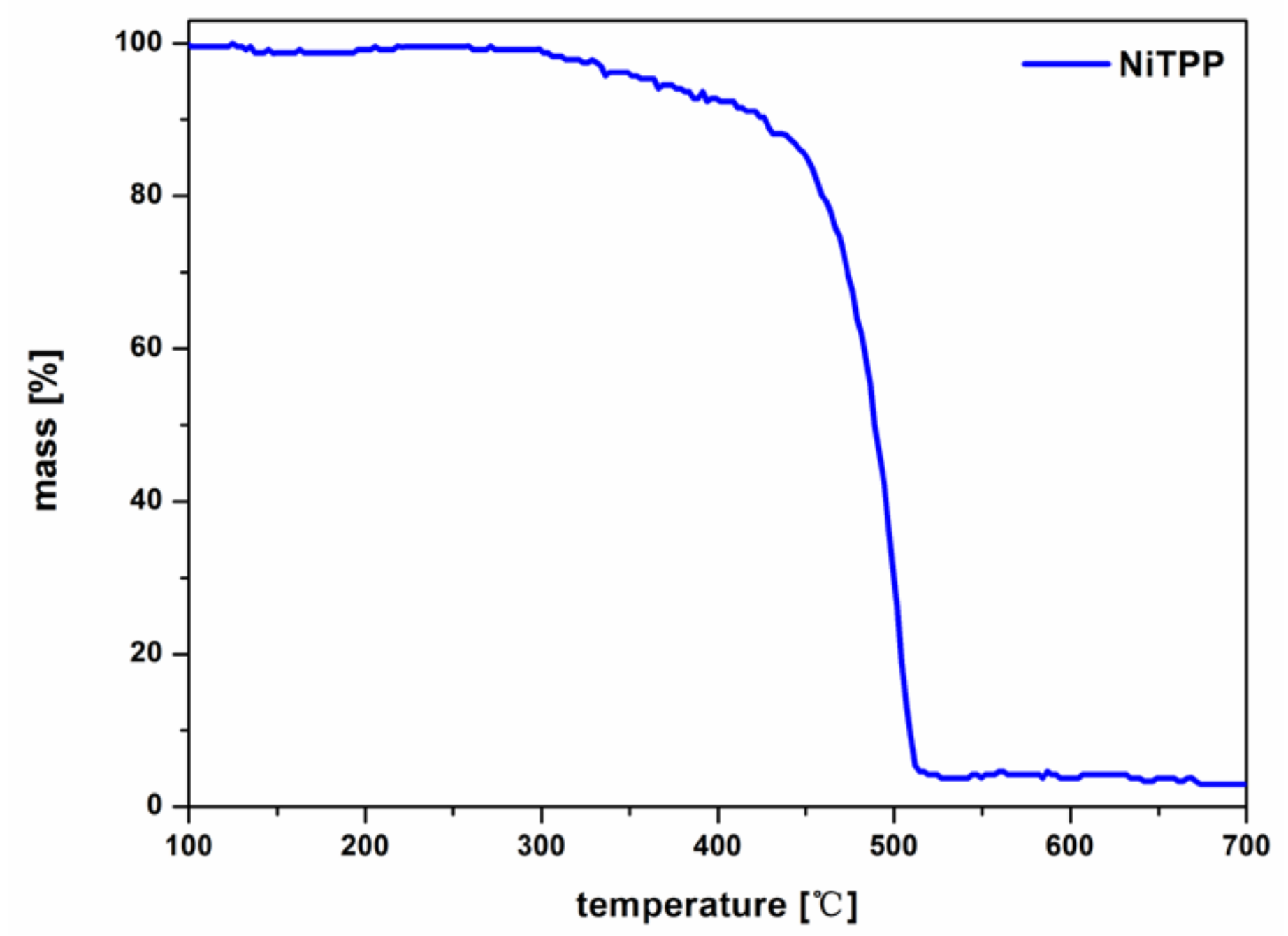

Figure S5. TGA curve of NiTPP.